\documentclass[a4paper,11pt,twoside]{scrartcl}
\usepackage{ILD}

\usepackage{xspace}

\usepackage{xpatch}
\makeatletter
\xpatchcmd\@HepConStyle
 {\edef\@upcode{\updefault}}
 {\ifdefined\shapedefault\edef\@upcode{\shapedefault}\else\edef\@upcode{\updefault}\fi}
 {}{}
\makeatother

\newcommand{\epem}{\ensuremath{\textrm{e}^{_+}\!\textrm{e}^{_-}}\xspace}

\newcommand{\geant}{\texttt{Geant4}\xspace}
\newcommand{\clupatra}{\texttt{Clupatra}\xspace}
\newcommand{\vzfinder}{\texttt{V0Finder}\xspace}
\newcommand{\kinkfinder}{\texttt{KinkFinder}\xspace}

\newcommand{\dmAH}[1]{$\Delta m = #1$\,GeV}
\newcommand{\Ma}[1]{$m = #1$\,GeV}

\title{Searching for long-lived particles with the ILD experiment}
\ildproc{PHYS}{2026}{006}

\date{March 19, 2026}

\addauthor{Jan Klamka}{\institute{a}}
\addauthor{Aleksander Filip \.Zarnecki}{\institute{b}}

\addinstitute{}{Faculty of Physics, University of Warsaw, Pasteura 5, 02-093 Warsaw, Poland}
\addinstitute{a}{jan.klamka@fuw.edu.pl}
\addinstitute{b}{filip.zarnecki@fuw.edu.pl}

\abstract{
Future \epem colliders provide a unique opportunity for long-lived particle (LLP) searches. We present a full simulation study of LLP searches using the International Large Detector (ILD), where a gaseous time projection chamber as the main tracking device provides excellent prospects for LLP searches. Signatures of displaced vertices and kinked tracks are explored. We study challenging final states involving both very soft displaced tracks and boosted, nearly collinear tracks. Backgrounds from beam-induced interactions and other Standard Model processes are considered. We present expected exclusion limits for a model-independent analyses, as well as for Higgs boson decays to LLPs, for a range of LLP lifetimes.
}

\titlecomment{Presented at the International Workshop on Future Linear
  Colliders (LCWS 2025),\\ Valencia, Spain, 20-24 October 2025.\\[5mm]
  This work was carried out in the framework of the ILD Concept Group}

\graphicspath{{./logos/} {./plots/} }

\begin{document}

\titlepage
\pagenumbering{arabic}\setcounter{page}{2}


\section{Introduction}

Despite the general consensus that there must be some physics phenomena Beyond the Standard Model (BSM), there is no clear indication how exactly they should manifest themselves in Nature. Many new-physics models predict the existence of the so-called long-lived particles (LLPs), which could travel macroscopic distances (from millimetres to metres, or more) after being produced, posing significant challenges to their detection. Future \epem Higgs factories, due to their clean experimental environment and, in some cases, trigger-less operation, seem very well-suited to directly search for such exotic states. In particular, the International Large Detector (ILD), an experiment proposed for operation at a future Higgs factory, offers great prospects for this kind of searches.

Results presented in this contribution are based on a recent PhD thesis \cite{Klamka:2026ccp} addressing prospects for detecting neutral and charged long-lived particles (LLPs) with  the ILD operating at the future \epem International Linear Collider (ILC) with $\sqrt{s}=250$\,GeV.
The analysis was based on a model-independent approach in which, instead of studying particular points in the parameter space of a specific model, benchmarks were selected based on the experimental signature and kinematic properties of the final state, allowing to test the detector capabilities and reconstruction tools. 
The study was performed using full \geant simulation for a set of benchmark scenarios. Multiple background sources were taken into account, both from low-$p_T$ beam-induced processes, as well as Standard Model interactions with high-$p_T$ final states.

\section{International Large Detector}

 The International Large Detector (ILD) was first proposed~\cite{LinearColliderILDConceptGroup-:2010nqx} as one of the experiments for the ILC and a design suitable for operation at a circular machine has been presented recently as well~\cite{ILD:2025yhd}.  The ILD is a general multipurpose detector, which relies on almost continuous tracking and highly granular calorimeters, and is optimised for the concept of particle-flow reconstruction~\cite{Thomson_2009}. 

 The large TPC as the main tracker is a very unique feature of the ILD.  It not only allows to detect more than 200 points on a charged particle's trajectory providing almost continuous tracking, but also enables particle identification with d$E$/d$x$ measurements. In the baseline design, the readout structure consists of 220 layers of \textit{pad rows} (or pad rings) with individual pads measuring $6\times1$\,mm$^{2}$.

Track reconstruction for the ILD is divided into several steps with track segments reconstructed separately in different sub-detectors and merged at the final stage. The \clupatra processor is used to find tracks in the TPC, while the \texttt{FullLDCTracking} processor matches and combines them with segments found in vertex and silicon detectors, and refits them globally~\cite{Gaede_2014}.

At the final stage of the ILD reconstruction chain, additional processors that operate on tracks are included.
\vzfinder attempts to identify and reconstruct the so-called V$^0$ particles, which is a collective term for long-lived neutral hadrons and converting photons. \kinkfinder is an analogous tool, but aimed at searching for ``kinked tracks''. In both cases, candidates that do not match any of the SM particles can be selected, which makes it useful for the BSM searches. 

\section{Search for neutral LLPs}

In the case of neutral LLPs, two classes of scenarios were analysed. One involved production of heavy scalar LLP, which decayed into SM gauge bosons and dark matter (DM). Scenarios with small mass difference between LLP and DM were selected, which provided non-pointing low-$p_T$ final states, difficult to distinguish from beam-induced backgrounds in a linear collider. The second class of benchmarks involved the production of a very light pseudoscalar, which can be produced in \epem collisions in association with a photon. Small LLP mass results in a large boost and high collimation of its decay products, bringing challenges for precise vertex reconstruction.

\begin{figure}[bt]
\includegraphics[width=0.49\textwidth]{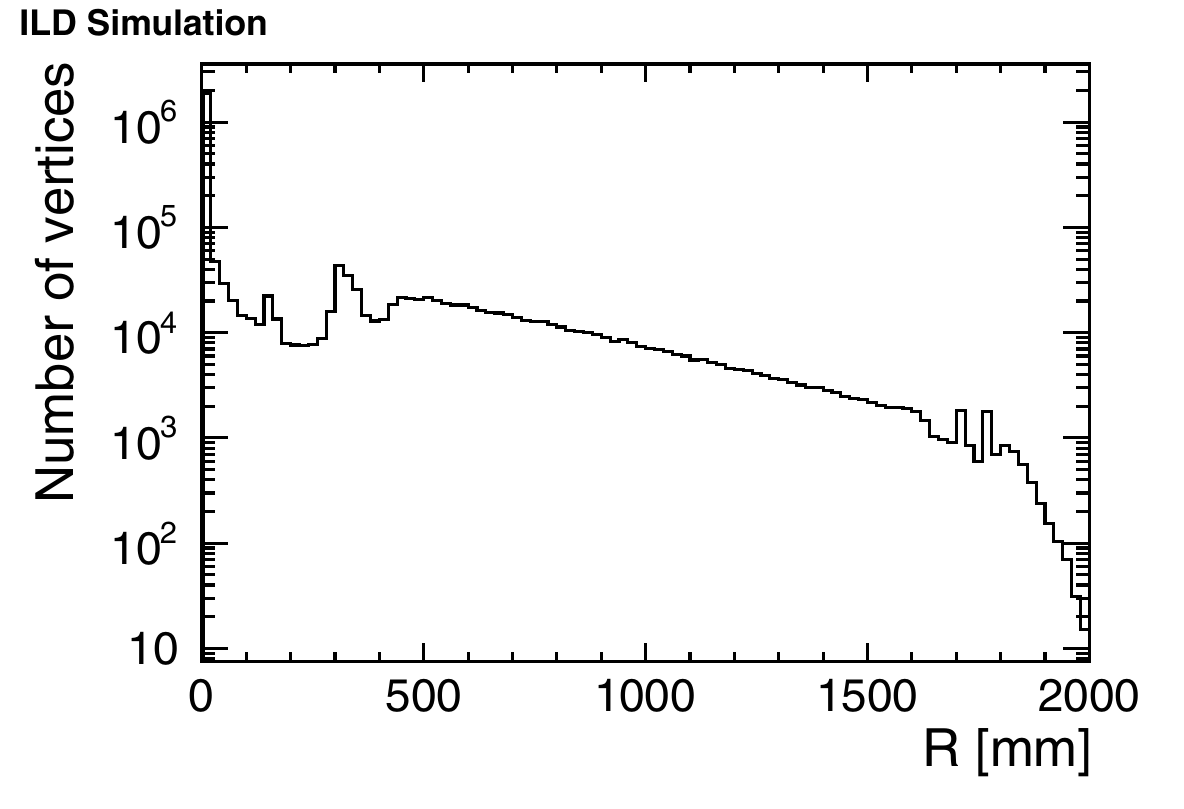}
\includegraphics[width=0.49\textwidth]{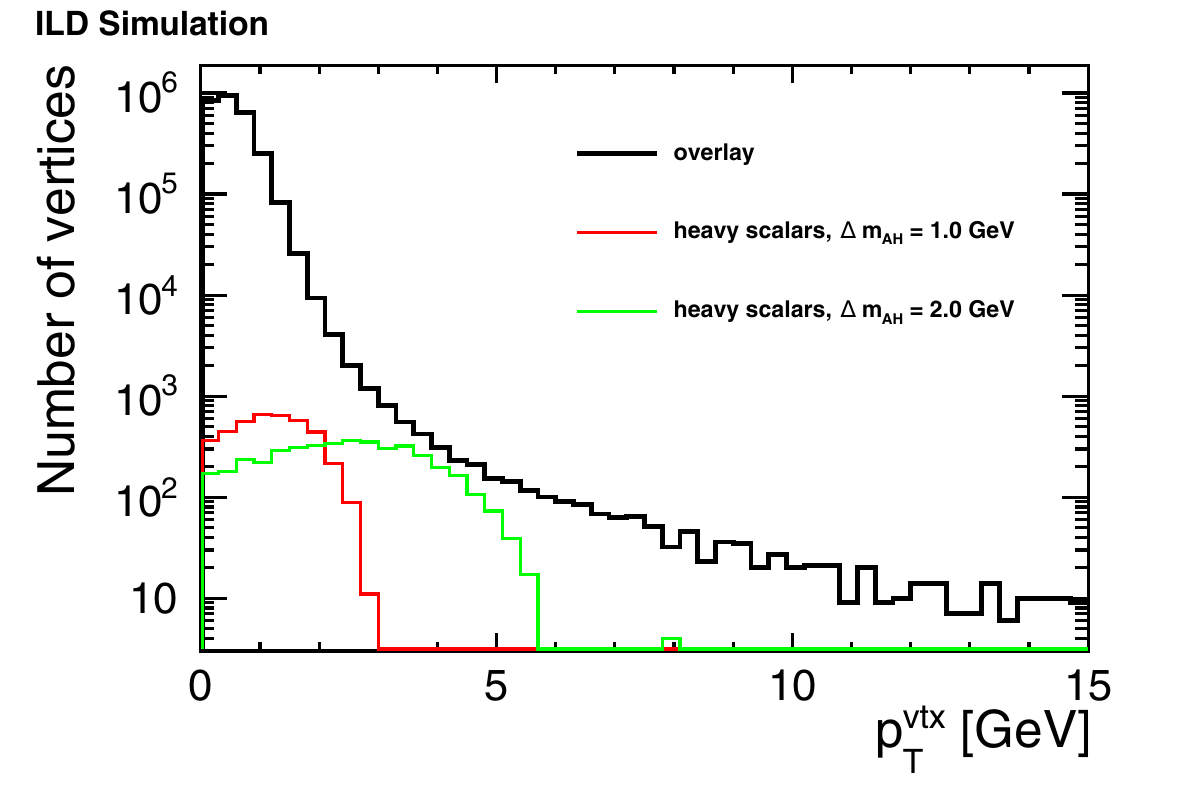}
 \caption{Left: Number of displaced vertices found in the overlay sample as a function of distance from the beam axis. Right: Total transverse momentum of tracks coming from a displaced vertex for the overlay (black) and scalar pair-production with mass difference between LLP and DM particle of $\Delta m=1$\,GeV (red) and $\Delta m=2$\,GeV (green). All histograms are normalized to the number of simulated MC events and correspond to no selection applied at all.  }
 \label{fig:background}
\end{figure}

The analysis was carried out using a vertex-finding algorithm designed specifically for this study. The search was performed considering only decays inside the TPC volume, with as few assumptions about the final state as possible -- in particular, final state objects other than the displaced vertex were ignored in the analysis. Soft, low-$p_T$ beam-induced (overlay) events were considered as a standalone background. Because of the high rate, they give sizable contribution, see Fig~\ref{fig:background}, and a dedicated selection procedure was proposed to suppress it. Hard, high-$p_T$ SM processes with hadronic jets and di-lepton events were also taken into account as background sources. Two working points were considered: standard model-independent selection and tight selection, which included additional cuts on event kinematics.

\begin{figure}[tb]
\includegraphics[width=0.49\textwidth]{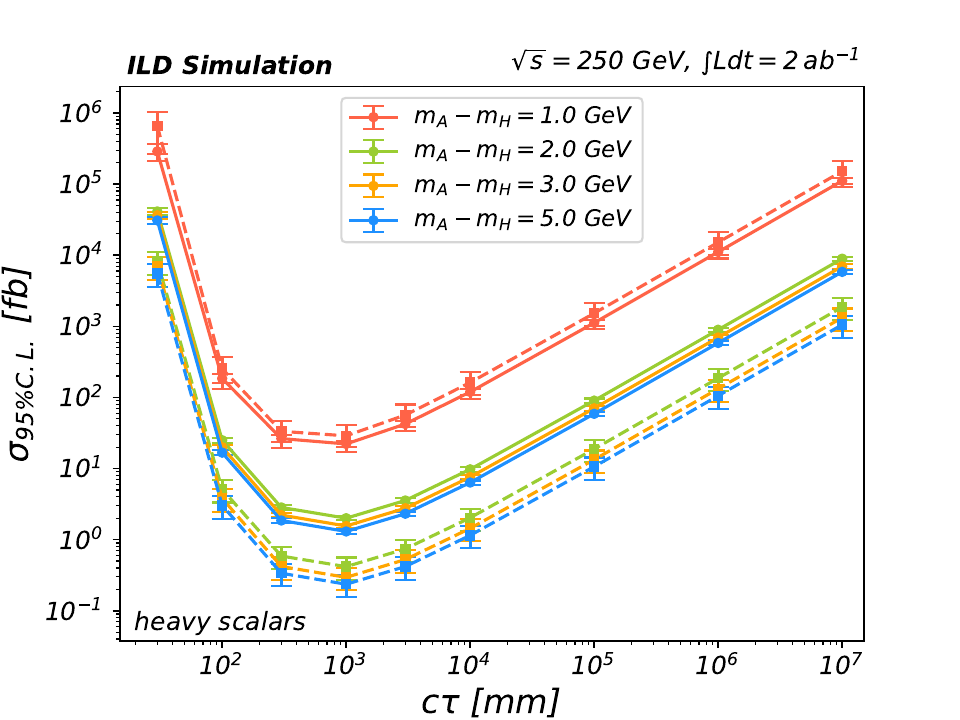}
\includegraphics[width=0.49\textwidth]{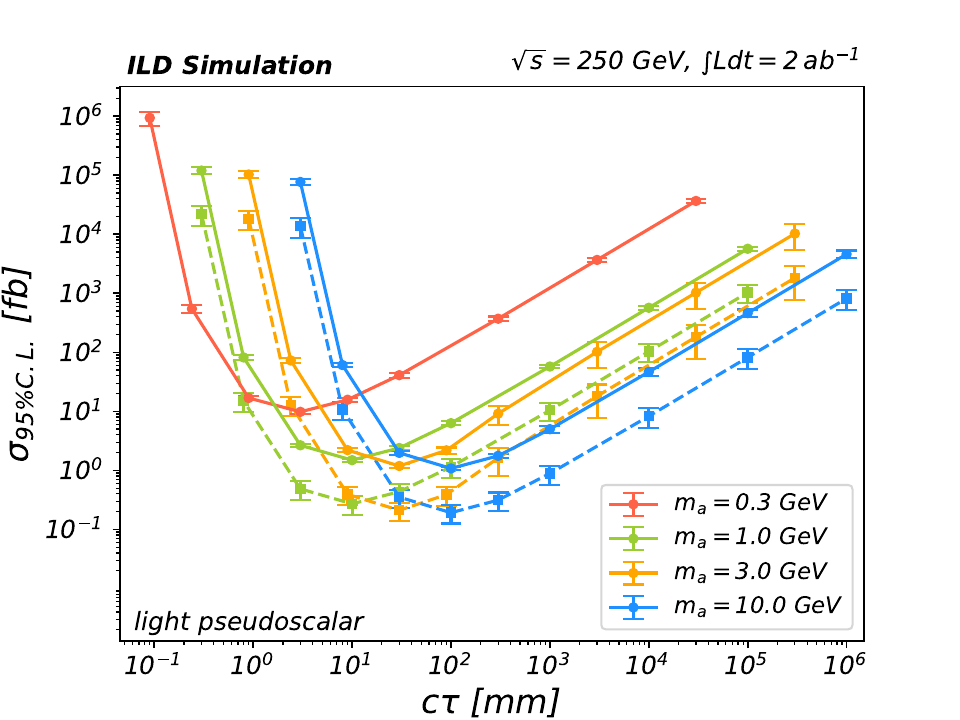}
 \caption{Expected 95\% C.L. upper limits on the signal production cross section for the considered benchmarks and different LLP mean decay lengths, for the scalar pair-production (left) and the light pseudoscalar production (right) at $\sqrt{s}=250$\,GeV. Solid lines corresponds to the standard selection and dashed lines to the tight set of cuts. The uncertainties are statistical. }
	 \label{fig:limits}
\end{figure}

Expected 95\% C.L. limits on the LLP production cross section were calculated for a range of lifetimes, using the obtained background levels and signal selection efficiencies. Results presented in Fig.~\ref{fig:limits} indicate that the ILD should be able to probe cross sections down to the level of femtobarns using standard selection for most of considered scenarios, and the tight selection provides an improvement by an order of magnitude. The most challenging benchmarks, production of the heavy scalar with the mass difference \dmAH{1} between LLP and DM, and the light long-lived pseudoscalar production with the mass \Ma{0.3}, could be accessed down to the level of 10\,fb using the standard model-independent selection~\cite{Klamka:2024gvd}.

\section{Exotic Higgs decays}

\begin{figure}[tb]
\includegraphics[width=0.49\textwidth]{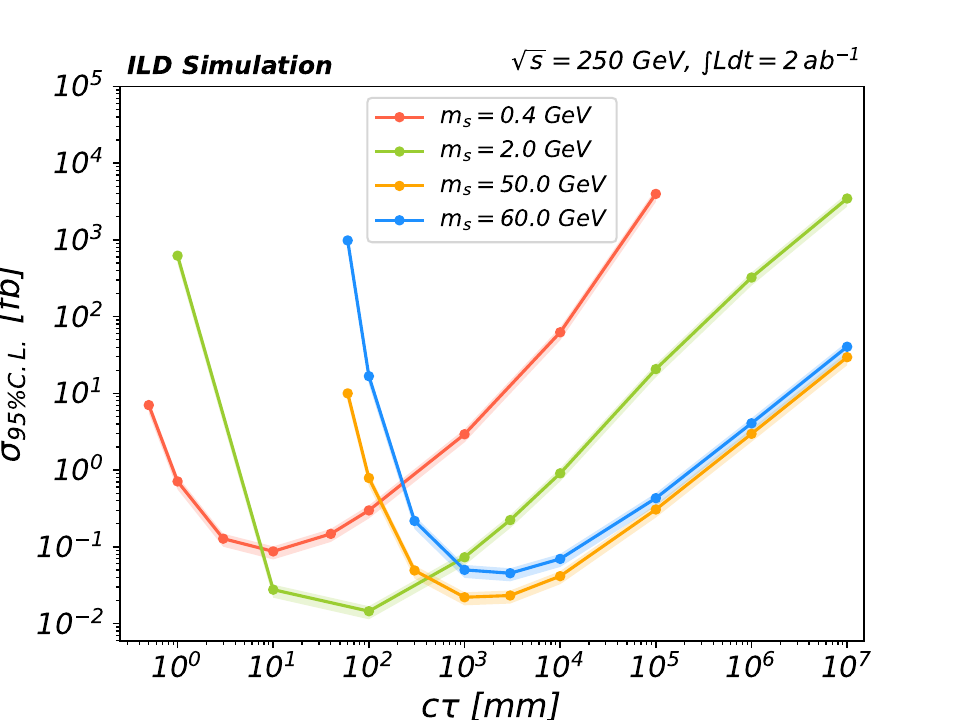}
\includegraphics[width=0.49\textwidth]{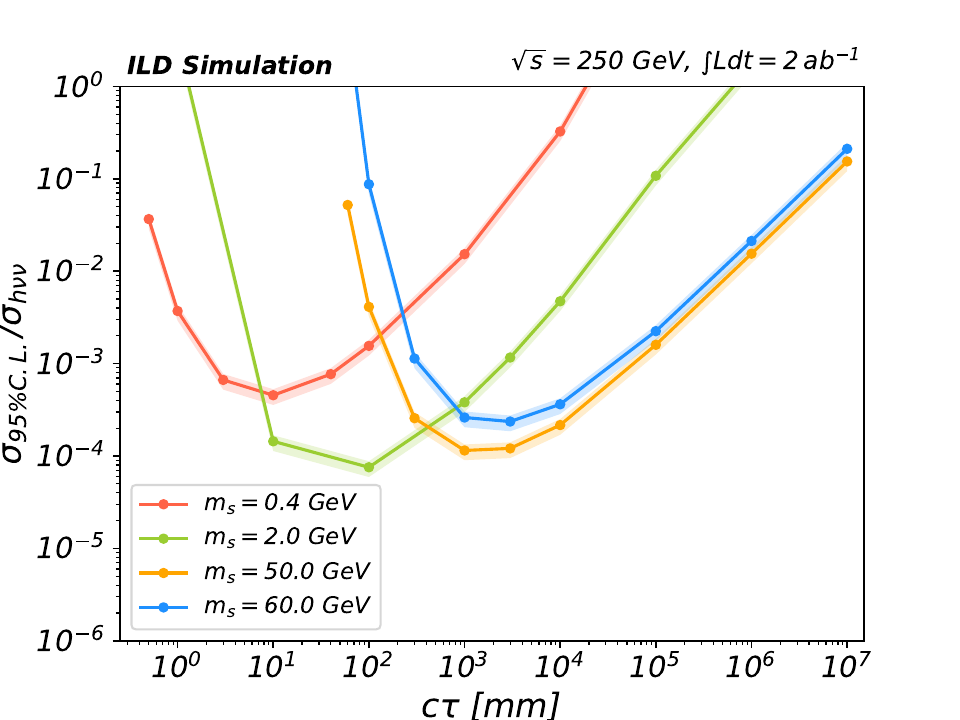}
 \caption{Expected 95\% C.L. upper limits on the signal production cross-section (left) and the branching ratio (right) for the considered benchmarks and different LLP mean decay lengths, for the Higgs decays to long-lived scalars at $\sqrt{s}=250$\,GeV. Shaded bands correspond to the limits calculated assuming $\pm1\,\sigma$ on the expected number of background events (see text for details). }
 \label{fig:higgs_llp_limits}
\end{figure}

Sensitivity to LLP production can be improved when optimising the search for a specific BSM scenario. This has been demonstrated in the search for exotic Higgs decays to LLPs, using the Higgsstrahlung production channel with the Z boson decays to neutrinos. The expected signature was at least one displaced vertex with no other activity inside the detector. Two scenarios with low LLP masses and two with high masses were considered. Additionally, requiring no high-$p_T$ prompt tracks in the event and using cuts on the total $p_T$ of tracks forming the vertex allowed to improve the reach by two orders of magnitude with respect to the previous analysis mentioned above.
Expected limits on the signal production cross-section  and the corresponding Higgs boson branching ratio limits are presented in Fig.~\ref{fig:higgs_llp_limits}.
The results on the Higgs branching ratio to dark scalars indicate that the ILD could provide an improvement with respect to current limits from the LHC in the regime of long lifetimes, depending on the LLP mass and BSM model, already at the first stage of ILC running. 

\section{Search for charged LLPs}

\begin{figure}[tb]
   \centerline{\includegraphics[width=0.5\textwidth]{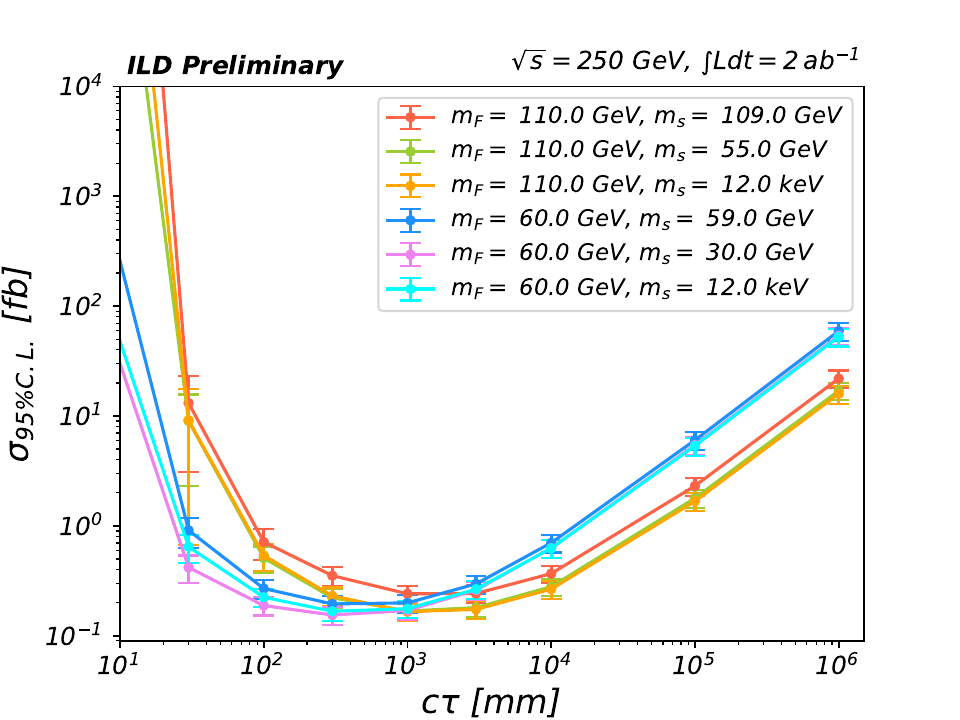}}
  \caption{Expected 95\% C.L. upper limits on the signal production cross section for all benchmarks considered in the charged LLP pair-production at $\sqrt{s}=250$\,GeV, shown as a function of the LLP proper decay length. The uncertainties are statistical. }
  \label{fig:fimp_limits}
\end{figure}

The charged LLP analysis was designed based on the experience gained in the search for displaced vertices.  A model predicting pair-production of long-lived dark fermions that decay into DM and SM leptons was used to generate six benchmark scenarios, with low and high masses of LLPs, and different mass differences between LLP and DM. The kinked track signature was considered in the analysis, and the search was carried out by optimising a tool designed for kink-finding which is available in the ILC software stack. A similar approach was taken to the model-independent analysis for displaced vertices, with the same background sources considered, but this time the search was performed in the entire detector volume. Extracted limits expected for the charged LLP production are shown in Fig.~\ref{fig:fimp_limits}. For all scenarios, cross sections below 1\,fb could be probed in the range of decay lengths from around 100\,mm to 10\,m, with the strongest limits reaching down to the level of 0.1\,fb. The results weakly depend on the mass difference between LLP and DM.

\section{Conclusions}

The results presented in this contribution demonstrate the great potential of the ILD experiment for LLP searches. The analysis framework developed allows to probe wide range of scenarios in a model-independent manner obtaining good sensitivity. Optimisation of selection criteria for a specific model provides further background reduction down to the level of single events resulting in orders of magnitude better sensitivity. The TPC plays a crucial role in the ILD discovery potential, which is confirmed not only by the results obtained in all presented analyses, but also by a direct comparison of the tracking acceptance with an alternative all-silicon ILD design. The results obtained could be even further improved by collecting more data at other stages of a linear collider operation.

\subsection*{Acknowledgements}

The work was supported by the National Science Centre (Poland) under OPUS research project no. 2021/43/B/ST2/01778. We would like to thank the LCC generator working group and the ILD software working group for providing the simulation and reconstruction tools and producing the Monte Carlo samples used in this study. This work has benefited from computing services provided by the ILC Virtual Organization, supported by the national resource providers of the EGI Federation and the Open Science GRID. 

\printbibliography

\end{document}